\documentclass[aps,prb,twocolumn,superscriptaddress,showpacs]{revtex4}
\usepackage{amssymb}
\usepackage{amsmath}
 
\usepackage[pdftex]{graphicx}

\usepackage{amsfonts}
\usepackage{bbold}
\usepackage{color}
 
\begin{document}
\title{On the correlation of light polarization in uncorrelated disordered magnetic media}
\author{M.A. Kozhaev}
\affiliation{Russian Quantum Center, Novaya 100A, Skolkovo, Moscow Region, 143025, Russia}
\affiliation{Prokhorov General Physics Institute RAS, Vavilov 38, Moscow, 119991, Russia}
\author{R.A. Niyazov}
\affiliation{NRC ``Kurchatov Institute", Petersburg Nuclear 
Physics Institute, Gatchina 188300, Russia}
\affiliation{Landau Institute for Theoretical Physics, Russian Academy of Sciences, Chernogolovka, Moscow oblast, 142432 Russia}
\author{V.I. Belotelov}
\affiliation{Russian Quantum Center, Novaya 100A, Skolkovo, Moscow Region, 143025, Russia}
\affiliation{Moscow State University, Leninskiye Gory 1, Moscow, Russia}
\date{\today}

\begin{abstract}

Light scattering in a magnetic medium with uncorrelated inclusions is theoretically studied in the approximation of ladder diagram. Correlation between polarizations of electromagnetic waves that are produced by infinitely-distant dipole source is considered. Here white noise disorder model with Gaussian distribution is taken into account. In such a medium the magneto-optical interaction leads to correlation between perpendicular light polarizations. Spatial field correlation matrix with nonzero nondiagonal elements is obtained in the first order on gyration.

\end{abstract}

\maketitle

\section{Introduction}

Recently there has been increased research interest in the polarization effects of light in scattering disordered media.~\cite{gasparian2013faraday, Vynck2014, dogariu2015electromagnetic} Light scattering was considered both in theory~\cite{akkermans1988theoretical, gorodnichev2009long, skipetrov2014optical, wang2015dynamic} and in experiment.~\cite{uchida2009large, strudley2014observation} This is mainly because of possibility for light localization.~\cite{lagendijk2009fifty} So far only weak localization has been observed~\cite{kaveh1986weak, jonckheere2000multiple} probably due to the vector nature of light.~\cite{skipetrov2014absence}

At the same time medium magnetization provides additional peculiarities in the scattering properties. For example, there is a photonic Hall effect in magnetic media.~\cite{rikken1996observation} Magnetic field could lead to light localization in a cold-atom gas.~\cite{skipetrov2015magnetic} There are multiple papers that study influence of magnetization on coherent backscattering. It was demonstrated that coherent backscattering decreases with magnetization increase due to the Faraday effect.~\cite{erbacher1993multiple} Albedo problem is considered in case of strong~\cite{Golubentsev} and small~\cite{Golubentsev1984} magnetooptical effects in comparison with disorder fluctuations.

One of the most popular systems under investigation is represented by magnetic metal--dielectric nanostructures.~\cite{armelles2013magnetoplasmonics} Such material leads to Faraday effect enhancement due to the scattering.~\cite{tkachuk2011plasmon, Belotelov2013, gevorkian2014plasmon} Media with natural optical activity as well as magneto--optical media were studied in paper.~\cite{MacKintoshJohn1988}

The classical model of light scattering ignores phase and polarization correlation on distances longer than the mean free path. The elastic scattering in macroscopic samples is considered as diffusion. But in many cases the interference between scattered light cannot be omitted. This fact highlights the need for well--described theoretical approach.

Key quantity for theoretical description of light propagation is a spatial field correlation matrix~\cite{LeonardMandel1995}. It is connected with observable quantities~\cite{Wolf1954}. The correlation matrix can be used to determine Stokes parameters~\cite{Goldstein2003,Ellis2004} and speckle pattern.~\cite{dogariu2015electromagnetic}

Theoretical description of Faraday rotation in scattering media was demonstrated in paper.~\cite{gasparian2013faraday} In the present paper we are going beyond this model. We propose a theoretical description for light scattering in infinite disordered magnetic media without absorption accurate to the ladder diagram.~\cite{rytov1978} We don't consider sub-leading correction of maximally-crossed diagrams which is important for the backscattering. 

Results are demonstrated up to the linear approximation on gyration. In uncorrelated nonmagnetic disordered medium only parallel field correlation components exist. In the magnetic medium the correlation matrix acquires nonzero nondiagonal elements. In other words the magnetooptical interaction leads to correlation of perpendicularly polarized components of scattered light at two different points. Nondiagonal part of the field correlation matrix is antisymmetric. Strict consideration of magnetooptical effect in scattering medium was demonstrated up to first order on gyration.

The paper is organized as follows. In Sec.~\ref{theory} we introduce the theoretical model of light propagation in a disordered magnetic medium. We briefly present the way of obtaining the electric field correlation matrix. Results obtained by this approach for different medium magnetization directions are shown in Sec.~\ref{results}. Sec.~\ref{conclusion} is devoted to discussion and conclusion.

\section{Theoretical model} \label{theory}

\subsection{Electric field in a disordered magnetic medium}

Electric field $E_i$ in a bulk medium, characterized by $\epsilon_{li}(\mathbf{r})$, is described by the Helmholtz equation.
\begin{equation}\label{maxeq}
(\partial_l \partial_i-\partial_k \partial_k \delta_{li} - \epsilon_{li} k_0^2) E_i=i \mu_0 \omega j_l,
\end{equation}
\noindent
where $\partial_k\equiv \partial / \partial r_k$, $\delta_{li}$ --- Kronecker symbol, $k_0=\omega/c$ is light wave vector in vacuum, $\omega$ is wave frequency, $c$ is speed of light, $j_l$ is electric current. Summation by repeated symbols is assumed here and elsewhere further.
In the simplest case of a non--magnetic and homogeneous medium the dielectric tensor is a diagonal matrix. In the magnetic medium the dielectric tensor takes the form:
\begin{equation}\label{dieltens}
\epsilon_{li}=\epsilon^0_l\delta_{li}- i e_{lik} g_k,
\end{equation}
\noindent
where $\epsilon^0_l$ is diagonal part of dielectric tensor, $e_{lik}$ is  Levi-Civita tensor, $g_k$ is gyration proportional related to the medium magnetization. Henceforth for simplicity, we assume that $\epsilon^0_l=1$. This case could be simply generalized by rescaling the relative dielectric tensor. 

The Helmhotz equation~\eqref{maxeq} can be solved by the Green function method. The Green function is the solution of equation~\eqref{maxeq} with $\delta$-function in the right side:
\begin{equation}
(\partial_l \partial_i-\partial_k \partial_k \delta_{li} - \epsilon_{li} k_0^2)  G_{ij}(\mathbf{r},\mathbf{r'})=\delta_{lj}\delta(\mathbf{r}-\mathbf{r'}).
\end{equation}
In that way electric field for arbitrary source can be expressed by:
\begin{equation}\label{elGreen}
E_i (\mathbf{r}) = i \mu_0 \omega \int G_{il}(\mathbf{r},\mathbf{r'}) j_l(\mathbf{r'}) d\mathbf{r'}.
\end{equation}

Retarded Green function for the magnetic medium in the reciprocal space $\mathbf{k}$ is given by:
\begin{equation} \label{greenf0}
\begin{aligned}
G^R_{il}(\mathbf{k})&=\sum_{\alpha=\pm1} P^\alpha_{il}(\hat{\mathbf{k}}) G^{R,\alpha}(\mathbf{k}),\\
 P^\alpha_{il}(\hat{\mathbf{k}})&=\frac 12\left(\delta_{il}-\hat{k}_i\hat{k}_l - i \alpha e_{ilj}\hat{k}_j\right), \\
 G^{R,\alpha}(\mathbf{k}) &=  \left[(1-\alpha \hat{\mathbf{k}} \cdot \mathbf{g}) k^2_0 - k^2 + i 0^+\right]^{-1},
\end{aligned}
\end{equation}
\noindent
where $\hat{\mathbf{k}} \equiv \mathbf{k}/k$ is unit vector along wave vector $\mathbf{k}$, $G^R(\mathbf{k})=\int G^R(\mathbf{r}) e^{-i \mathbf{k} \mathbf{r}} d\mathbf{r}$ is retarded Green function in the reciprocal space, $\alpha$ is connected to right ($\alpha=+1$) and left ($\alpha=-1$) circular polarizations of scattered light. Such structure of the Green function with projectors $\mathbf{P}^{1 (-1)}$ on right (left) circular polarization is caused by the magnetooptical interaction. The sign before $0^+$ in the denominator of $G^{R,\alpha}(\mathbf{k})$ is selected to obtain only outgoing waves.

Let us consider a magnetic medium with disorder. The disorder is given by a small fluctuation $\delta \epsilon(\mathbf{r})$ of the dielectric tensor i.e. we add a diagonal term to $\epsilon_{li}$ in~\eqref{dieltens}. The disorder model is a white noise:
\begin{equation}\label{whitenoise}
\langle \delta \epsilon(\mathbf{r})\rangle=0, \quad \langle\delta \epsilon(\mathbf{r}) \delta \epsilon(\mathbf{r'})\rangle = \frac{6 \pi}{ l k_0^4} \delta(\mathbf{r}-\mathbf{r'}),
\end{equation}
\noindent
where $\langle\ldots\rangle$ means averaging by the Gaussian disorder distribution and $l$ is elastic mean free path of the medium. 

In equation~\eqref{greenf0} the approximate retarded Green function is shown. Indeed, accurate retarted Green function contains of longitudinal part which is connected to the nontransversality~\cite{Stark1997}. In work~\cite{M.LandiDeglInnocenti2004} it was demonstrated that in nonmagnetic medium with optically anisotropic inclusions the longitudinal part of electric field is proportional to the $O(\delta\epsilon(\mathbf{r}))$. In our case if light propagates along the gyration vector $(\mathbf{g} \parallel \mathbf{k})$ then its longitudinal electric field component is absent.~\cite{Eroglu2010} However if light propagates perpendicular to the gyration vector $(\mathbf{g} \perp \mathbf{k})$ then the longitudinal component gives only $O(g^2)$ terms in Green function.

Consideration of multiple scattered events on sparse inclusions in Bourret approximation~\cite{bharucha2014probabilistic} (i.e. in the first--order on $\zeta=1/k_0l$) results in substitution of $0^+$ by $k_0/l$ in the Green function:~\cite{Golubentsev,Golubentsev1984,MacKintoshJohn1988}
\begin{equation}
 G^{R,\alpha}(\mathbf{k}) =  \left[(1-\alpha \hat{\mathbf{k}} \cdot \mathbf{g}) k^2_0 - k^2 + i k_0/l\right]^{-1}.
\end{equation}

\subsection{Field correlation matrix}

The field correlation matrix $W_{kl}(\mathbf{r},\mathbf{r'})$ is defined by:
\begin{equation}
W_{kl}(\mathbf{r},\mathbf{r'})=\langle E_k(\mathbf{r}) E_l^*(\mathbf{r'})\rangle .
\end{equation}
It can be described by Bethe-Salpeter equation which in the ladder approximation~\cite{Vynck2014} reads:
\begin{equation}\label{BSeq}
\begin{aligned}
\langle E_k(\mathbf{r}) E_l^*(\mathbf{r'})\rangle = &\langle E_k(\mathbf{r}) \rangle \langle E_l^*(\mathbf{r'})\rangle \\ 
&+ k_0^4 \int \langle G_{km}(\mathbf{r},\mathbf{r_1}) \rangle \langle G^*_{ln}(\mathbf{r'},\mathbf{r'_1})\rangle \\
&\times  \langle \delta \epsilon(\mathbf{r_1}) \delta \epsilon(\mathbf{r'_1}) \rangle\langle E_m(\mathbf{r_1}) E_n^*(\mathbf{r_1'}) \rangle d\mathbf{r_1} d \mathbf{r_1'}.
\end{aligned}
\end{equation}

\begin{figure}[h] \begin{center}
 \includegraphics[width=1\linewidth]{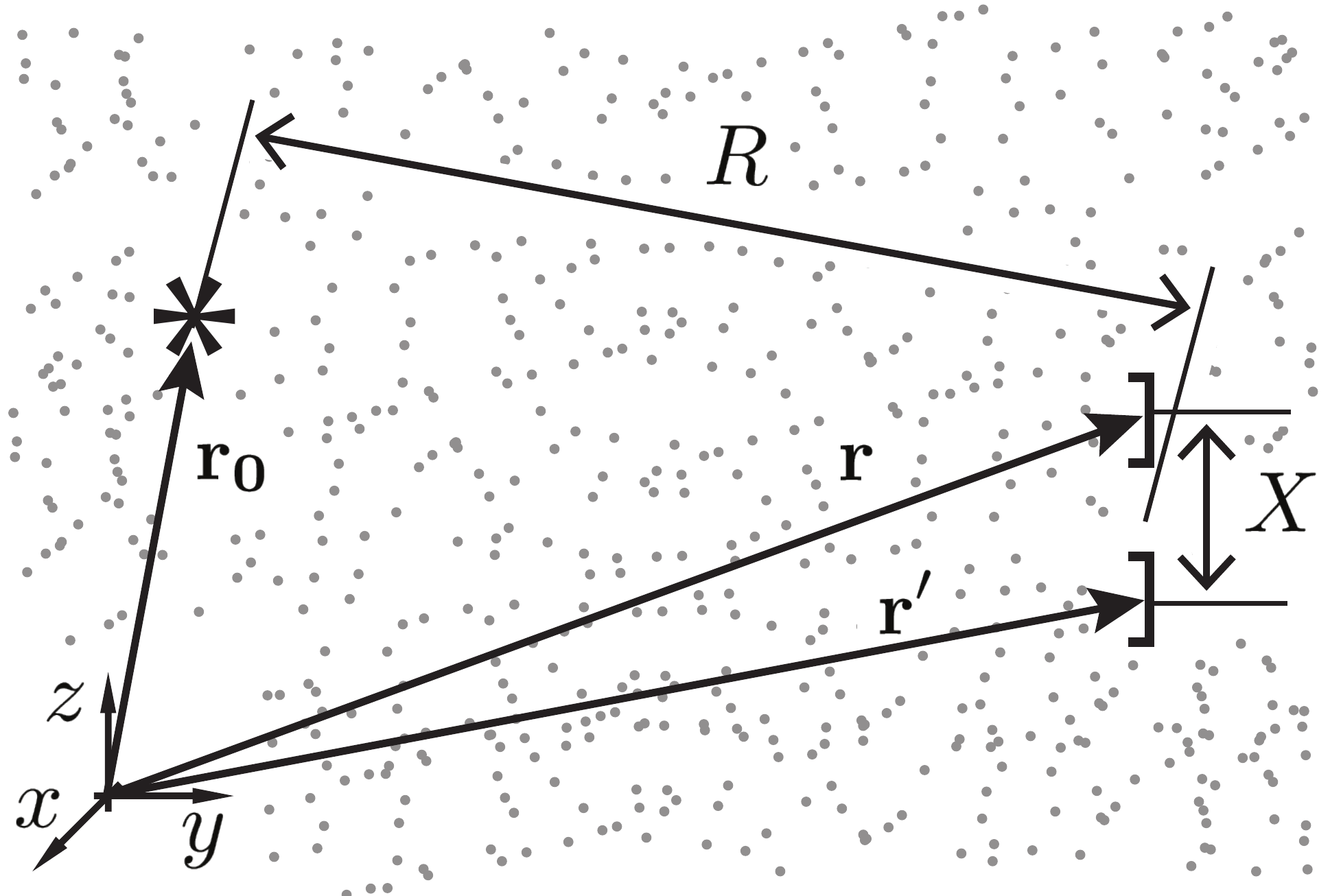}
\caption{Observation scheme of light fields correlation in infinite medium with disordered inclusions. The light source at point $\mathbf{r_0}$ is at a distance $R$ from the detectors middle point. The detectors at points $\mathbf{r}$ and $\mathbf{r'}$ are separated by a distance $X$. Hereinafter $\mathbf{X}\parallel z$.}  
\label{fig:detectors}
 \end{center}  \end{figure}

We assume that the light source is a point dipole with a dipole moment $p=1/\mu_0 \omega^2$ located at $\mathbf{r_0}$:
\begin{equation}
j_l (\mathbf{r},\mathbf{r_0})= - i \omega p_l \delta (\mathbf{r}-\mathbf{r_0}).
\end{equation}
Equation~\eqref{BSeq} can be solved in reciprocal space by the Fourier transform on variables $\mathbf{X}=\mathbf{r}-\mathbf{r'}$ and $\mathbf{R}= (\mathbf{r}+\mathbf{r'})/2 -\mathbf{r_0}$ with dual variables in the reciprocal space, correspondingly, $\mathbf{q}$ and $\mathbf{K}$ (see Fig.~\ref{fig:detectors}).

The first step is to solve the Bethe-Salpeter equation with $\mathbf{X}=0$:
\begin{equation}\label{BS0}
D_{ijkl}(\mathbf{K})=S_{ijkl}(\mathbf{K})+\frac{6\pi}{l} S_{ijmn} (\mathbf{K}) D_{mnkl}(\mathbf{K}),
\end{equation}
\noindent
where
\begin{equation}
\begin{aligned}
 S_{ijkl}(\mathbf{K})=&\int \langle G_{ik}(\mathbf{q}+\mathbf{K}/2) \rangle \langle G^*_{jl}(\mathbf{q}-\mathbf{K}/2) \rangle \frac{d\mathbf{q}}{(2\pi)^3}, \\
 D_{ijkl}(\mathbf{K})= &\int \langle G_{ik} (\mathbf{q}+\mathbf{K}/2)  G^*_{jl}(\mathbf{q}-\mathbf{K}/2) \rangle \frac{d\mathbf{q}}{(2\pi)^3}.
 \end{aligned}
\end{equation}
 We can find $D_{ijkl}$ by obtaining eigenvalues and eigenvectors $S_{ijkl}$:
 \begin{equation}\label{DfromS}
\begin{aligned}
\frac{6 \pi}{l} S_{ijkl} (\mathbf{K}) &= \sum_{p=1}^{9} \lambda_p |ij\rangle_p \langle kl|_p \Rightarrow \\
D_{ijkl}(\mathbf{K})&= \sum_{p=1}^{9} \frac{l}{6 \pi} \frac{\lambda_p}{1-\lambda_p}|ij\rangle_p \langle kl|_p .
\end{aligned}
\end{equation}
 Next step is to solve equation~\eqref{BSeq} for $\mathbf{X} \neq 0$:
 \begin{equation}
 \begin{aligned}
Q_{ijkl}(\mathbf{q},\mathbf{K}) = & F^{\alpha \beta}_{ijkl}(\hat{\mathbf{q}},\mathbf{K})\\
& \times \langle G^\alpha(\mathbf{q}+\mathbf{K}/2) G^{*\beta}(\mathbf{q}-\mathbf{K}/2) \rangle, \\ 
 F^{\alpha \beta}_{ijkl} (\hat{\mathbf{q}},\mathbf{K})=& P^\alpha_{ik}(\hat{\mathbf{q}}) P^{*\beta}_{jl}( \hat{\mathbf{q}})\\
 & + \frac{6 \pi}{l} P^\alpha_{im}( \hat{\mathbf{q}}) P^{*\beta}_{jn}(\hat{\mathbf{q}}) D_{mnkl}(\mathbf{K}),
\end{aligned}
\end{equation}
where $\hat{\mathbf{q}} \equiv \mathbf{q}/q$ is unit vector along wave vector $\mathbf{q}$.

Without loss of generality, we suppose that the source is a point dipole oriented along $z$-axis. In such a way the correlation matrix in reciprocal space is:
\begin{equation}\label{qtow}
W_{ij}(\mathbf{q},\mathbf{K})=Q_{ij33}(\mathbf{q},\mathbf{K}).
\end{equation}
Finally we have to find the inverse Fourier transform of~\eqref{qtow}. The Fourier transform of $W_{ij}(\mathbf{q},\mathbf{K})$ made only on $\mathbf{X}$, $W_{ij}(\mathbf{R},\mathbf{q})$, is a local density matrix of the radiation with wave vector $\mathbf{q}$.~\cite{Golubentsev} Thereby we introduce quantity $I_{ij}(\mathbf{R},\hat{\mathbf{q}})$ by integration $W_{ij}(\mathbf{R},\mathbf{q})$ on the absolute value of $\mathbf{q}$:
\begin{equation}
I_{ij}(\mathbf{R},\hat{\mathbf{q}}) = 2 \int q^2 W_{ij}(\mathbf{R}, q \hat{\mathbf{q}})dq.
\end{equation}
It is proportional to the light intensity.~\cite{Stephen1986}

\section{Results}\label{results}

Further we will consider the following approximation: 
\begin{equation}\label{eq:approximation}
g^2\ll (Kl)^2 \ll g \ll \zeta \ll 1
\end{equation}
It corresponds to the situation when the light source is located far from the detectors $R \gg l$, elastic scattering plays the main role and the magnetooptical effect can be considered as a small correction. Linear on gyration contribution comes from eigenvectors of $\mathbf{S}$. As was discovered earlier~\cite{MacKintoshJohn1988} the eigenvalues contain only $O(g^2)$ terms. The eigenvectors compution is rather complicated task. Only recently eigenvectors without gyration were found up to $O\left((Kl)^2\right)$.~\cite{Vynck2014} For analytic analysis of linear on gyration contribution we take into account only largest $Kl$--terms that are $O\left(1/(Kl)^2\right)$. Further details about approximate solution of Bethe-Salpeter equation could be found in the Appendix~\ref{bssolving}.

Let us assume that observation points are along $z$-axis: $\mathbf{X} \parallel \mathbf{z}$. We consider two magnetization directions: $\mathbf{g_\parallel}$ ($\mathbf{g} \parallel \mathbf{z}$) and $\mathbf{g_\perp}$ ($\mathbf{g}\parallel\mathbf{y}$). Normalized quantity $\tilde{\mathbf{W}}(\mathbf{X})=\mathbf{W}(\mathbf{R},\mathbf{X})/W(\mathbf{R})$, where $ W(\mathbf{R}) =\left.  \mbox{Tr}[\mathbf{W}(\mathbf{R},\mathbf{X}) \right|_{\mathbf{X}=0}]= \frac{k_0 \zeta }{8 \pi ^2 R}$, does not depend on the distance from the light source.

\begin{figure}[h] \begin{center}
\includegraphics[width=1\linewidth]{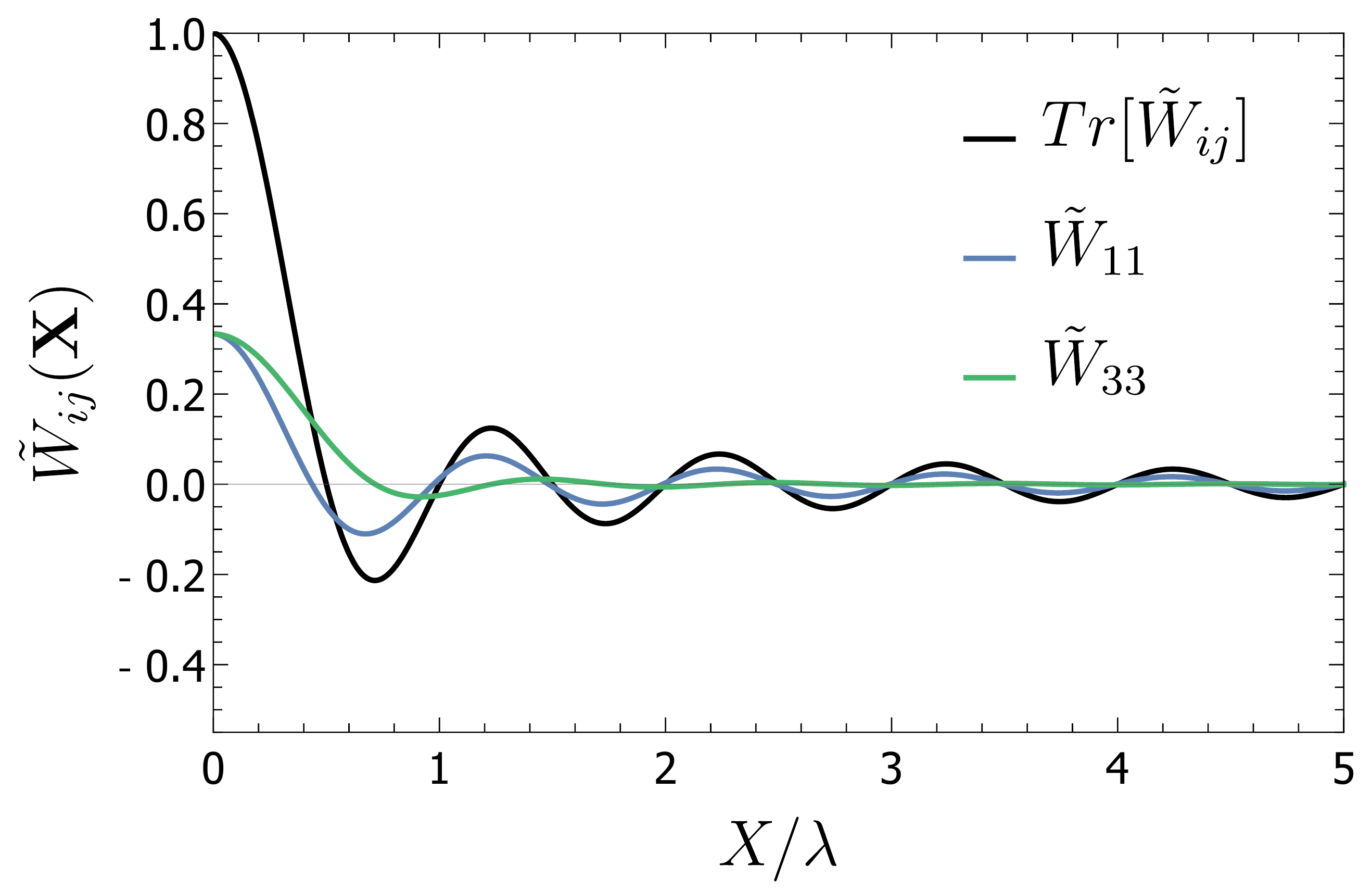}
\caption{Dependence of the diagonal normalized field correlation matrix components on the distance between the detectors. $\mathbf{X}$ is oriented along the $z$-axis. $W_{ij}$ shows the correlation between $E_i$ and $E_j$ components on the detectors.  We assume that the mean free path $l=20 \lambda$. The distance between the detectors is normalized on wavelength.}
\label{fig:speckle1}
\end{center}
\end{figure}

\begin{figure}[h] \begin{center}
\includegraphics[width=1\linewidth]{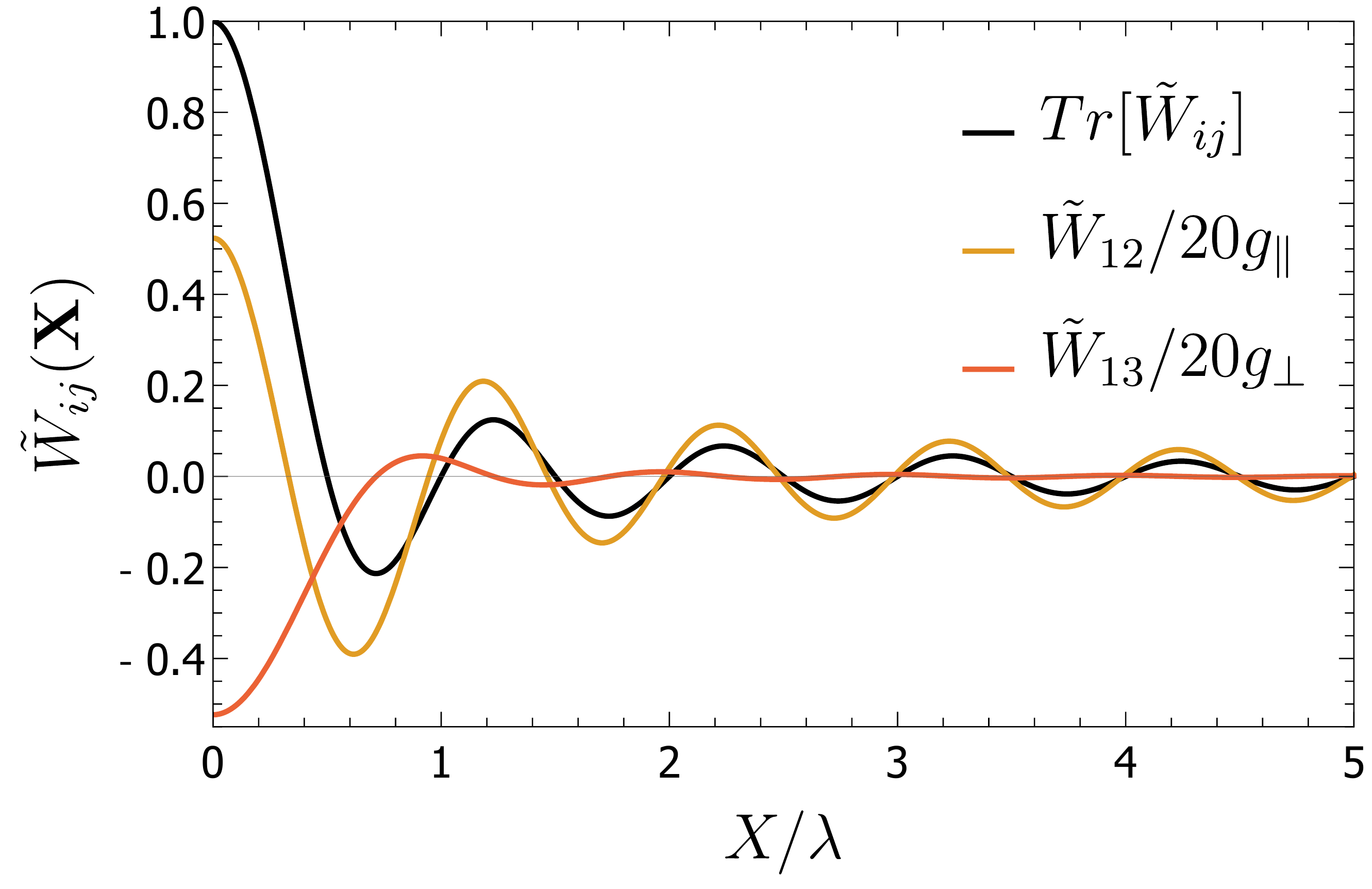}
\caption{Dependence of the nondiagonal normalized field correlation matrix components on the distance between detectors. $\mathbf{X}$ is oriented along the $z$-axis. $W_{ij}$ shows the correlation between $E_i$ and $E_j$ components on the detectors. Yellow line corresponds to the normalized nonzero element of the correlation matrix divided by $20 g_\parallel$ when gyration is directed parallel to $\mathbf{X}$. Red line corresponds to the normalized nonzero element of the correlation matrix divided by $20 g_\perp$ when gyration is directed perpendicular to $\mathbf{X}$. We assume that the mean free path $l=20 \lambda$. The distance between the detectors is normalized on wavelength.}  
\label{fig:speckle2}
 \end{center}  \end{figure}

We obtain the same diagonal elements for both gyration orientations:
\begin{equation}\label{winX}
\begin{aligned}
\tilde{W}&_{11}(X)=\tilde{W}_{22}(X)=
 \\
&\frac{\left((k_0 X)^2-1\right) \sin{k_0 X}+k_0 X \cos{k_0 X}}{2 (k_0 X)^3}- \\&\zeta \frac{  \left((k_0 X)^2-2\right) (k_0 X \sin{k_0 X}+\cos{k_0 X})+2}{4 (k_0 X)^3},\\
\tilde{W}&_{33}(X)=
\frac{\sin{k_0 X}-k_0 X \cos{k_0 X}}{(k_0 X)^3} \\&+\zeta\frac{  \left((k_0 X)^2-2\right) \cos{k_0 X}-2 k_0 X \sin{k_0 X}+2}{2 (k_0 X)^3}.
\end{aligned}
\end{equation}
This diagonal elements have no magnetooptical contribution and coincide with the result with zero gyration~\cite{Vynck2014} (see Fig.~\ref{fig:speckle1}). However there are nonzero nondiagonal elements proportional to $g/\zeta$ (Fig.~\ref{fig:speckle2}). For $\mathbf{g_\parallel}$ they are:
\begin{multline}
\label{winX3}
\tilde{W}_{12}(X)=-\tilde{W}_{21}(X)= \\
\frac{g_\parallel}{\zeta} \frac{\left((k_0 X)^2-2\right) \sin{k_0 X}+2 k_0 X \cos{k_0 X}}{4 (k_0 X)^3 }.
\end{multline}
While for $\mathbf{g_\perp}$:
\begin{multline}\label{winX2}
\tilde{W}_{13}(X)=-\tilde{W}_{31}(X)= \\
\frac{g_\perp}{\zeta} \frac{ (k_0  X \cos {k_0 X}- \sin{k_0 X})}{4(k_0  X)^3}.
\end{multline}

If $\mathbf{g}\parallel\mathbf{X}$ only two nondiagonal components $\tilde{W}_{12}=-\tilde{W}_{21}$ are nonzero. However, for $\mathbf{g}\perp\mathbf{X}$ only $\tilde{W}_{13}=-\tilde{W}_{31}$ are nonzero and decays faster.

 \section{Discussion and Conclusion}\label{conclusion}

To conclude, we have theoretically investigated light scattering in a magnetic medium with uncorrelated inclusions. The spatial field correlation matrix with ladder approximation is used for light propagation description, so only two-particle interactions are considered. The approximation of the distant light source, when the distance from the source $R$ is much larger than the elastic scattering mean free path, is taken into account up to the order $O\left(1/(Kl)^2\right)$. The magnetooptical interaction is described by terms linear in gyration. The distance between the detectors $X$ is assumed much smaller than the distance to the source.

Here we studied the correlation matrix dependency on direction and amplitude of the sample magnetization. Explicit calculations of the eigenvalues and eigenvectors were performed in degenerate perturbation theory as it was previously suggested for nonmagnetic medium in.~\cite{Vynck2014} The first--order magnetooptical effect correction was found. Nonzero values of the nondiagonal field--correlation matrix components are found. Such components are responsible for correlation of the perpendicular polarizations of the scattered light between different points. Influence of the magnetization direction 
 on correlation of light polarization is demonstrated.

This work extends the understanding of light behavior in magnetic scattering medium. The correlation between perpendicular polarizations of light opens the possibility to develop time--reversal--noninvariant systems based on magnetooptical effects enhanced by scattering.~\cite{tkachuk2011plasmon, Belotelov2013, gevorkian2014plasmon} Rigorous theory of light scattering in the presence of a magnetic field may lead to better understanding of ferrofluids magnetooptics.~\cite{Brojabasi2015} Moreover, it might be interesting for study of light scattering in magnetoplasmonic structures.~\cite{Khokhlov2015, Belotelov2014, Krutyanskiy2013}

Further investigation in this area may go in different directions. First of all, this theory can be applied to various forms of scattering media: scatterers with anisotropic dielectric tensor or inhomogeneous gyration. Secondly instead of infinite medium it is possible to consider half--space or slab geometries. Additional development allows to obtain better theoretical model up to orders of $O\left((Kl)^2\right)$ and $O\left(g^2\right)$. 
Also it is possible to extend theory to account backscattered light which requires consideration of maximally crossed diagrams.

\acknowledgements

We are grateful to Andrey Kalish and Ilya Pusenkov for numerous useful discussions. This work was supported in part by the Russian Foundation for Basic Research (grant N 16-02-01065). The work of R.N. was supported by the Russian Science Foundation (grant No. 16-42-01035).

\appendix
\section{Solving Bethe-Salpeter equation}\label{bssolving}

First, we have to compute the $\mathbf{S}$-tensor from equation~\eqref{BS0}. With the aim to compute a result in the analytic form we have to choose certain limited orders in small parameters of our problem. We can compute the $\mathbf{S}$-tensor at least in orders $O\left( (Kl)^2 \right)$ and $O\left( g^2 \right)$ but eigenvectors getting is a rather difficult computational task. This can be done in the degenerate perturbation theory. 

Main contribution in the small parameter $K l$ is $(1/Kl)^2$. It rises only from one eigenvalue of the $\mathbf{S}$-tensor: $\lambda = 1 - (Kl)^2/3$. This eigenvalue has only second order gyration correction. We neglect such term assuming that $g^2 \ll (Kl)^2$. But $\mathbf{S}$-tensor eigenvectors have first order in gyration corrections. For their accounting we assume that $(Kl)^2\ll g$. Thus we have the following relations of small parameters of our problem: $g^2\ll (Kl)^2 \ll g \ll \zeta \ll 1$.

Detailed computation of $S_{ijkl}$ with $O((1/Kl)^2)$ terms but without gyration can be founded in Vynck et al~\cite{Vynck2014}. To obtain $\mathbf{S}$ with nonzero gyration we compute integral on module $q$ by residue theorem. After that we can expand $\mathbf{S}$ because $g \ll \zeta$ and take angular integral. To compute eigenvectors for the first order of gyration we use the degenerate perturbation theory. Since $\mathbf{S}$ with gyration is a non-symmetric matrix right and left eigenvectors must be distinguished. We use only eigenvectors which correspond to unit eigenvalue and coincide. For arbitrary gyration orientation:
\begin{equation}
 |ij\rangle_{1}=\langle ij|_{1} =\frac{\sqrt{3}}{3} \delta_{ij}+\frac{\sqrt{3}}{6} e_{ijk} g_k.
\end{equation}

After computation $\mathbf{D}$ from~\eqref{DfromS} we can obtain $\mathbf{Q}$. We are interested in $1/(Kl)^2$ term which is located only in $\mathbf{D}$. Consequently, we can neglect first term of $\mathbf{F}$ and $K$-dependence of the Green function.

From $\mathbf{D}$ we can find $\mathbf{W}$~\eqref{qtow}:
\begin{equation}\label{eq:cohmat}
\begin{aligned}
W_{ij} (\mathbf{K},\mathbf{q})= & \frac{1}{(Kl)^2} \sqrt 3 |mn\rangle_1 P^\alpha_{im}(\hat{\mathbf{q}}) P^{*\beta}_{jn}( \hat{\mathbf{q}}) \\
 & \times \langle G^\alpha (\mathbf{q}) \rangle \langle G^{*\beta}(\mathbf{q}) \rangle.
\end{aligned}
\end{equation}
Summation by repeated symbols is assumed here. In nondiagonal elements there are two distinct contributions of the orders $O(g/ \zeta)$ and $O(g)$. We neglect
 $g$-order term due to relation $\zeta \ll 1$. 
Inverse Fourier transform of $K$-depended part of the correlation matrix proportional to $1/R$. Fourier transform from the $q$-space to $X$-space leads to~\eqref{winX}--\eqref{winX2}.

\end{document}